\documentclass[pdflatex,sn-mathphys]{sn-jnl}

\jyear{2022}
\usepackage{siunitx}
\usepackage{subcaption,graphicx}
\usepackage{multirow}
\usepackage{gensymb}
\usepackage{overpic}

\theoremstyle{thmstyleone}

\theoremstyle{thmstyletwo}

\theoremstyle{thmstylethree}

\begin{document}
\title[Thermal Transport in Twisted Bilayer Graphene]{Thermal Transport in Twisted Bilayer Graphene: An Equilibrium Molecular Dynamics Study}

\author*[1]{\fnm{} \sur{David}}\email{david18001@mail.unpad.ac.id}
\author[1,2]{\fnm{Ferry} \sur{Faizal}}\email{ferry.faizal@unpad.ac.id}
\author[1,2]{\fnm{I Made} \sur{Joni}}\email{imadejoni@phys.unpad.ac.id}

\affil*[1]{\orgdiv{Department of Physics}, \orgname{Faculty of Mathematics and Natural Sciences}, \orgaddress{\street{Universitas Padjadjaran}, \city{Jl. Raya Bandung-Sumedang Km 21 Jatinangor}, \postcode{45363}, \state{Jawa Barat}}}
\affil[2]{\orgdiv{Functional Nano Powder University Center of Excellence (Finder U-CoE)}, \orgname{Universitas Padjadjaran}, \orgaddress{\city{Jl. Raya Bandung-Sumedang Km 21 Jatinangor}, \postcode{45363}, \state{Jawa Barat}}}

\abstract{Twisted bilayer graphene (tBLG) is two graphene layers placed on top of each other with a twist angle, making it has tunable thermal properties. In this paper, we report an analysis of thermal conductivity ($\kappa$), phonon density of states, and specific heat capacity of tBLG with various twist angles over a range of temperatures using equilibrium molecular dynamics simulations based on the Green--Kubo method. Simulation shows that stacking and twisting graphene layers lead to a decrease in the thermal conductivity, with the highest $\kappa$  at around room temperature owned by the tBLG with a twist angle of 3.89\textdegree followed by 16.43\textdegree and 4.41\textdegree. We also perform quantum correction to the simulation results to show the process of increasing thermal conductivity at low temperatures.}


\maketitle
\section{Introduction}\label{sec1}
Due to the rapid development of technology, the demand for compact devices is increasing. This phenomenon causes the industry to manufacture smaller electronic devices from time to time. Shrinking the electronics' size will increase the power density and make them heat up easily. The higher the temperature during operation makes electronic devices perform poorly, become unstable, and have a shorter lifetime.

Recently, research on thermal material has been increasing, motivated by the heat dissipation problem in electronics and the curiosity to understand heat conduction at the nanoscale. This researches mainly focus on thermal conductivity, a measurement of how well a material can transfer or conduct heat. Materials with high thermal conductivity can be used as heat removal and thermal interface materials (TIM), and Graphene is one of them.

Graphene is a carbon-based 2D material with unique characteristics. At room temperature, its thermal conductivity can reach 5300 W/mK, making it has the largest thermal conductivity among all known materials\cite{balandin2008superior,katsnelson2020crystal}. Giving some modifications like stacking and twisting on graphene layers will give a significant impact on its thermal properties\cite{wang2020frank, zhang2020molecular, zhang2022controllable}. Currently, many researchers have a high interest in investigating the heat conduction process of twisted bilayer graphene (tBLG), where two graphene layers are placed on top of each other with a twist angle. By giving this modification, the thermal properties become tunable and angle-dependent.

To investigate more about graphene's thermal properties, we can look closer at its energy carrier, whereas in carbon materials the dominant contribution comes from the phonons\cite{hone2001phonons}. Phonons are quasiparticles that describe the vibrational motions of a lattice. They are divided into acoustic and optical modes, depending on the directions of the lattice vibration. Low-frequency phonons are referred to as acoustic phonons, where atoms vibrate coherently in the same direction. The optical phonons are made by the out-of-phase lattice vibration with a frequency much higher than the acoustic phonons. Phonon behavior is dictated through scattering mechanisms, with the most common phonon–phonon scattering is the Normal processes (or N processes) and Umklapp processes (or U processes)\cite{shin2020advanced}. Normal processes occur when two phonons interact and then combine into a new phonon, or one phonon splits into two new phonons, with both energy and momentum being conserved. Umklapp processes happen when two phonons meet and combine to form a new phonon with an amplitude larger than the Brillouin zone\cite{boer2018phonon}. In this case, the new combined phonon would have a different direction from the vector direction of the vector addition of the initial two phonons. Therefore, the energy is conserved in the U process, but the momentum is not. It is well known that the Umklapp processes dominate at high temperatures and create resistance to heat flow, while the Normal processes only redistribute phonons. Phonon contribution to thermal conductivity is described by Klemens-Callaway\cite{klemens1958lattice} model:

\begin{equation}
	\kappa=\frac{1}{3} v \Lambda C_{V}
\end{equation}

\noindent where $\Lambda$ is the phonon mean free path and $C_{V}$ is the specific heat capacity, and both are related to the phonon frequency\cite{dong2020role, tian2011importance, toberer2012advances}. Therefore, the key to understanding the thermal conductivity of materials is to understand the phonon characteristic. Still, the exploration of these thermal properties is time-consuming and expensive if carried out experimentally.

Molecular dynamics (MD) simulation is a well-known method for investigating material properties, including thermal conductivity. To calculate thermal conductivity using MD, there are two methods that can be used: equilibrium molecular dynamics (EMD) and non-equilibrium molecular dynamics (NEMD). The EMD method uses the Green-Kubo formulation with the system's temperature being kept constant during the simulation. On the other hand, thermal conductivity calculation using NEMD is based on Fourier's law and involves the temperature gradient of the system. Compared to the NEMD, the EMD has more advantage for calculating thermal conductivity in small systems with periodic boundary conditions because the calculation results are not affected by the size of the system\cite{vasilev2022prediction,khadem2013comparison}.

There are related research has been reported before. Wang et al.\cite{wang2017thermal} had investigated the thermal properties of tBLG but only focused on the large twist angle (0\textdegree, 30\textdegree, and 60\textdegree). Their research did not explain the thermal conductivity of tBLG at a small angle, which currently attracts many researchers since Cao\cite{cao2018unconventional} found superconductivity in this system. Nie et al.\cite{nie2019interlayer} have measured the thermal conductivity of tBLG and twisted multilayer graphene using NEMD. Still, according to several studies, measurements with NEMD resulted in a smaller value than the experimental result, and size effects in NEMD are more severe than EMD\cite{khan2017thermal}. So in this study, we aim to investigate the thermal conductivity of tBLG with the variation of small angle by using equilibrium molecular dynamics simulations.

\section{Methods}\label{sec2}
In this study, we have carried out equilibrium molecular dynamics (EMD) simulations using the Large-scale Atomic/Molecular Massively Parallel Simulator (LAMMPS)\cite{LAMMPS} for investigating the thermal properties of twisted bilayer graphene (TBLG). The twist angles considered here are 0\degree (AA-BLG), 3.89\degree, 4.41\degree, and 16.43\degree. The three angles were chosen since their supercell can construct tBLG system with similar cell length, to minimize the size effect on thermal conductivity calculation. The supercell structures are generated using ASE\cite{larsen2017atomic} \texttt{flatgraphene} module so we can obtain the coordinates of each constituent atom on x, y, and z axes.

Each tBLG can be represented by the index (m,n)\cite{uchida2014atomic}, where m and n are non-negative integers with m $\le$ n which indicates BLG replication along the x and y directions. This index can give other information related to the size of the twist angle ($\theta$), the width of the unit cell ($L_{cell}$) and the number of atoms ($N_{atom}$) in a supercell through the following equation:

\begin{equation}
	\cos \theta=\frac{n^{2}+4 n m+m^{2}}{2\left(n^{2}+n m+m^{2}\right)}
\end{equation}

\begin{equation}
	L_{cell}=d\sqrt{3\left(n^{2}+m n+m^{2}\right)}
\end{equation}

\begin{equation}
	N_{atom}=4\left(n^{2}+nm+n^{2}\right)
\end{equation}

Besides the tBLG structure, simulations were also performed on single-layer graphene (SLG) and bilayer graphene (BLG), to see thermal conductivity value in untwisted graphene. The SLG structure in this study was obtained by modifying the structure of tBLG 3.89$\degree$ by removing the upper layer. While the BLG structure is obtained by duplicating the SLG structure on the z-axis so that the BLG structure is obtained with the AA configuration. The structure of commensurate lattice is shown in the following figure:
\begin{figure}[H]
	\centering
	\includegraphics[width=12cm]{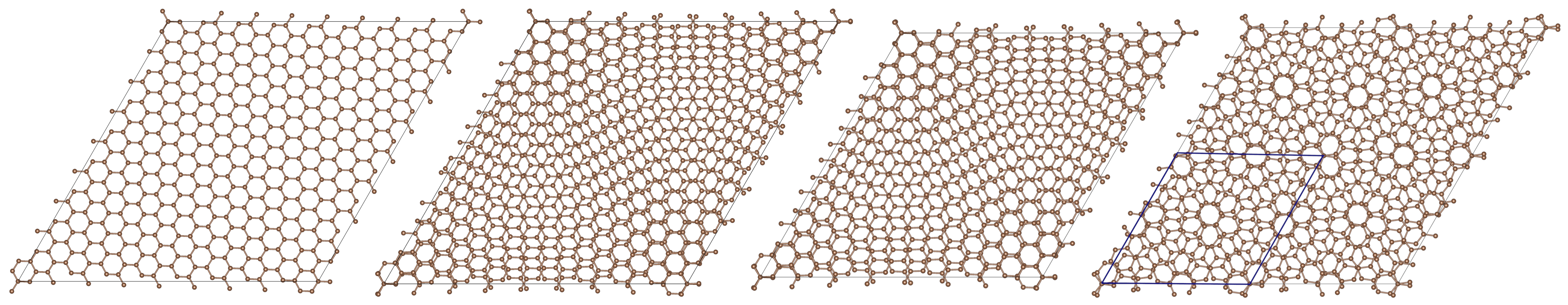}
	\caption{Structure of Twisted Bilayer Graphene}
\end{figure}

\noindent The 16.43$\degree$ tBLG structure has a length of 17.10 \AA \space with 196 atoms in one supercell. However, in this study this structure is duplicated twice on the x and y axes to make it has approximately similar length with other systems, in order to minimize the size effect in the calculation and comparison of thermal conductivity. All graphene system we have calculated in this work are summarized in Table \ref{tab:graphene}.

\begin{table}[ht!]
	\centering
	\caption{A list of graphene system calculated in this work}
	\begin{tabular}{|l|l|l|l|l|l|}
		\hline
		\textbf{System} & \textbf{m, n} & \textbf{\begin{tabular}[c]{@{}l@{}} $\theta$ (\degree)\end{tabular}} & \textbf{\begin{tabular}[c]{@{}l@{}}Duplication\\ ($x \times y \times z$)\end{tabular}} & \textbf{\begin{tabular}[c]{@{}l@{}}$L_{cell}$ (\AA)\end{tabular}} & \textbf{\begin{tabular}[c]{@{}l@{}}$N_{atom}$\end{tabular}} \\ \hline
		SLG             & -             & -                                                                    & 1 x 1 x 1                                                                              & 35.98                                                                                  & 434                                                                              \\ \hline
		BLG             & -             & 0                                                                    & 1 x 1 x 1                                                                              & 35.98                                                                                  & 868                                                                              \\ \hline
		tBLG            & 9, 8          & 3.89                                                                 & 1 x 1 x 1                                                                              & 35.98                                                                                  & 868                                                                              \\ \hline
		tBLG            & 8, 7          & 4.41                                                                 & 1 x 1 x 1                                                                              & 31.75                                                                                  & 676                                                                              \\ \hline
		tBLG            & 5, 3          & 16.43                                                                & 2 x 2 x 1                                                                              & 34.20                                                                                  & 784                                                                              \\ \hline
	\end{tabular}
	\label{tab:graphene}
\end{table}

Adaptive intermolecular reactive empirical bond order (AIREBO) potential has been considered to model the interactions between C–C bonds:

\begin{equation}
	U_{ij}=\frac{1}{2} \sum_{i} \sum_{j \neq i}\left[U_{i j}^{\mathrm{REBO}}+U_{i j}^{\mathrm{LJ}}+\sum_{k \neq i, j} \sum_{l \neq i, j, k} U_{k i j l}^{\mathrm{TORSION}}\right]
\end{equation}

\noindent where is $U^{\mathrm{REBO}}$ represents the covalent interactions between pairs of atoms according to the Reactive Bond Order (REBO) potential developed by Brenner\cite{brenner2002second}, $U^{\mathrm{LJ}}$ portion represents long-distance intermolecular interactions between atoms that are not directly bonded. This section is based on the Lennard-Jones 12-6 potential. The $U_{kijl}^{\mathrm{TORSION}}$ term is used to describe various dihedral angle preferences in hydrocarbon configurations. This part was considered not to be taken into account in the graphene study because of the absence of C-H bonds. The corresponding parameters for the AIREBO potential are provided in Ref. \cite{stuart2000reactive}.

In our study, periodic boundary conditions were applied in the x and y-direction. A time step of 1 fs was used for the simulations. Equations of atomic motion were integrated using a Velocity-Verlet integrator. Energy minimization is obtained using the fast inertial relaxation engine (FIRE)\cite{bitzek2006structural} algorithm. The system is equilibrated using NVT ensemble for $1\cdot10^{5}$ fs to achieve a well-optimized system before thermal conductivity calculation is started. The thermal conductivity as a function of temperature along the i-th (i = x, y, z) direction can be computed by Green-Kubo formalism:

\begin{equation}
	\kappa_{i}=\frac{1}{V k_{B} T^{2}} \int_{0}^{\infty}\left\langle J_{i}(0) J_{i}(t)\right\rangle d t
\end{equation}

\noindent where $\kappa$ is the thermal conductivity, $k_B$ is the Boltzmann constant, $T$ is the system temperature, V is the system volume, $\tau$ is the required correlation time for the reasonable enough decay of HCACF and $\langle\vec{J}(0)\cdot\vec{J}(t)\rangle$ is the ensemble averaging term. The heat flux is given by:

\begin{equation}
	{J}=\sum_{i} E_{i} {v}_{i}+\sum_{i>j}\left[{F}_{i j} \cdot\left({v}_{i}+{v}_{j}\right)\right] {r}_{i j}
\end{equation}

\noindent where $v_i$ is the velocity of atom i, $F_{ij}$ is the force on an atom i exerted by its neighboring atom j, and $r_{ij}$ is the relative position vector. Here, the total energy associated with the i-th atom, $E_i$, is expressed by

\begin{equation}
	E_{i}=\frac{1}{2} m_{i} {v}_{i}^{2}+\frac{1}{2} \sum_{i \neq j} U_{i j}
\end{equation}

\noindent where $m_{i}$ is the atomic mass, and $U_{ij}$ is the potential function. To see the accuracy of the simulation results, the error can be measured by the following equation:

\begin{equation}
	SE = \frac{\sqrt{\frac{1}{N} \sum_{i=1}^{N}\left(\kappa_{i}-\bar{\kappa}\right)^{2}} }{\sqrt{N}}
\end{equation}

\begin{equation}
	RRMSE = \frac{\sqrt{\frac{1}{N} \sum_{i=1}^{N}\left(\kappa_{i}-\bar{\kappa}\right)^{2}} }{\bar{\kappa}}\times100\%
\end{equation}

\noindent where SE is the standard error and RRMSE is the relative root-mean-square error, or referred to as relative error. The standard error values are displayed directly on the graph of thermal conductivity vs. temperature as error bars, while the relative errors are shown on a separate graph for deeper analysis.

After equilibration, time series data for the heat current was collected under NVE for $9\cdot10^{5}$ fs. The thermal conductivity value in the x, y, and z-direction are evaluated from the heat current autocorrelation function (HCACF). The integral of
the HCACF was evaluated using the trapezoid rule.

Quantum correction for thermal conductivity calculations of classical MD is required because quantum effects in classical MD approximation below the Debye temperature ($T_d$) are neglected\cite{khan2015equilibrium, kaviany2014heat}. To investigate the thermal conductivity at low temperatures, it is necessary to find the specific heat capacity using PDOS\cite{islam2019anomalous}. The PDOS describes the number of different states per unit volume that phonons can occupy at a particular energy level. It is obtained by taking the fast Fourier transform of the velocity autocorrelation function (VACF) after equilibrium is achieved. Therefore, for the phonon with frequency $\omega$, the PDOS can be calculated as

\begin{equation} \label{eq:PDOS}
	G(\omega)=\int \frac{<v_{i}\left(0\right) \cdot v_{i}\left(t\right)>}{<v_{i}\left(0\right) \cdot v_{i}\left(0)>\right.} e^{-2 \pi i \omega t} dt
\end{equation}

\noindent where $v_i(t_0)$ is the velocity of the i-th atom at time $t_0$, $<>$ means an average over a set of initial times $t_0$.

The quantum correction factor of the heat capacity which is the ratio of $C_v$ according to Bose-Einstein statistics (because phonons are bosons) and classical $C_v$ ($3Nk_B$), as shown in the following equation:

\begin{equation}
	f_{Q}=\frac{C_v(T)}{3Nk_B}=\frac{\displaystyle\int\limits_{0}^{\infty}\frac{u^2e^u}{{(e^u-1)}^2}G(\omega)d\omega}{\displaystyle\int\limits_{0}^{\infty}G(\omega)d\omega}
	\label{eq:QCF}
\end{equation}

\noindent where $u=\frac{\hbar \omega}{k_{B} T}$. The corrected thermal conductivity value is obtained by this equation:

\begin{equation}
	\kappa_{QC}=f_{Q} \cdot \kappa_{MD}
\end{equation}

To see the accuracy of the simulation, the calculation results of heat capacitance are compared with experimental data. The relationship between heat capacitance from simulation and experiment can be shown through the correlation coefficient which can be obtained by equation below

\begin{equation}
	\mathrm{r}_{\mathrm{C}_{\mathrm{VS}} \mathrm{C}_{\mathrm{VE}}}=\frac{\sum \mathrm{xy}}{\sqrt{ \left(\sum \mathrm{x}^{2} \mathrm{y}^{2}\right)}}
\end{equation}

\noindent with $\mathrm{x}=\left(\mathrm{C}_{\mathrm{VSi}}-\overline{\mathrm{C_{VS}}}\right)$ and $\mathrm{y} =\left(\mathrm{C}_{\mathrm{VEi}}-\overline{\mathrm{C_{VE}}}\right)$.

\section{Results}\label{sec3}

\subsection{Equilibration}
Choosing the timestep value is very important in Molecular Dynamics because it determines the simulation stability. The stable simulation is indicated by the energy kept in constant value. But, since the simulation is based on discrete equations, the energy value can’t be perfectly constant, so there will be small fluctuations over time.

\begin{figure}[H]
	\centering
	\includegraphics[width=10cm]{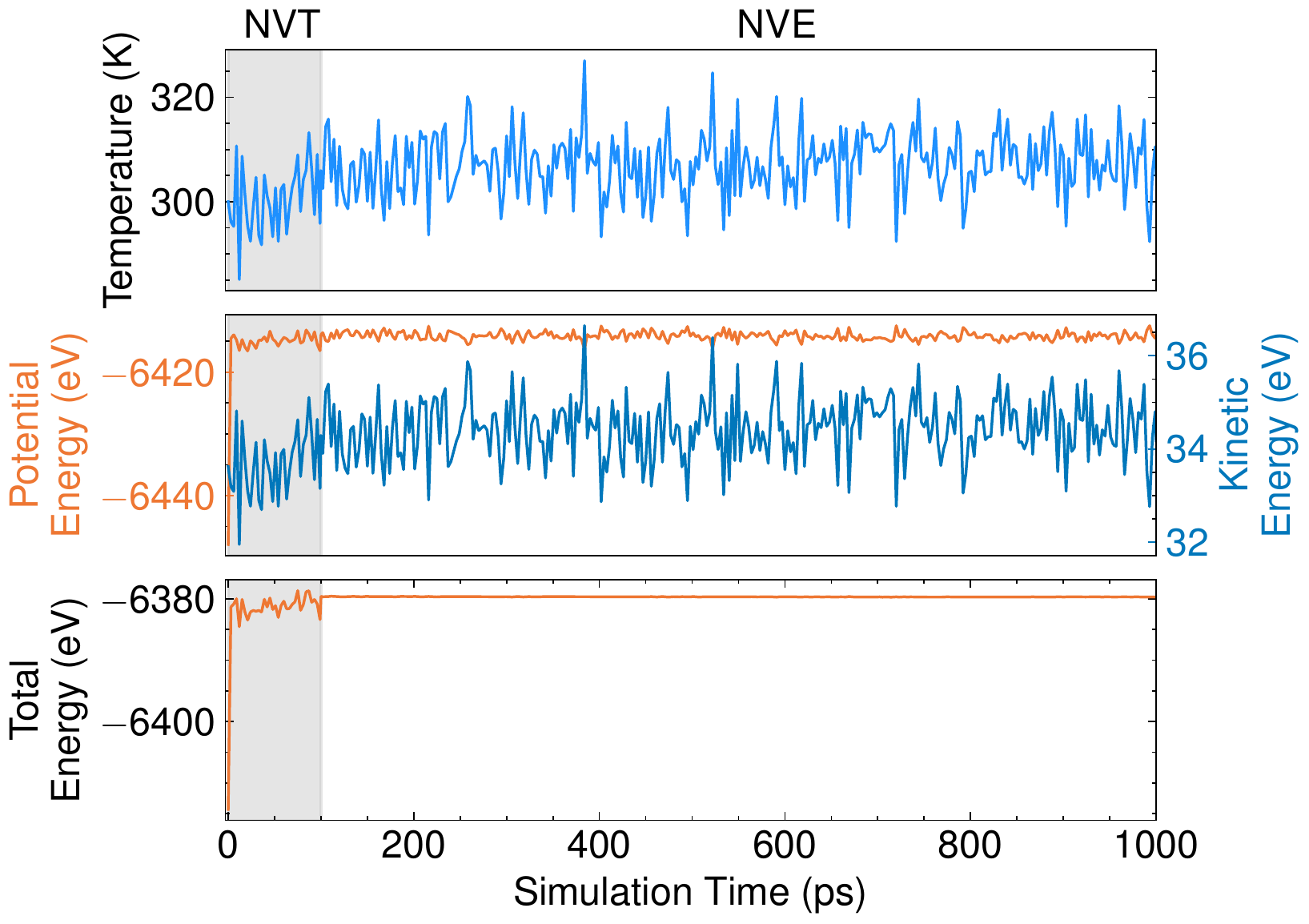}
	\caption{Energy and temperature during the equilibrium processes (tBLG = 3.89\textdegree, T = 300 K)}
	\label{fig:equilibriumpro}
\end{figure}

Figure \ref{fig:equilibriumpro} shows how the temperature and energy of tBLG 3.89\textdegree at 300 K change during the equilibration process. At the beginning of the simulation, the total energy increases and then fluctuates, until it becomes constant and reaches equilibrium at 100 ps. At this stage, when potential energy increases, the kinetic energy is decreasing, and vice versa, making the sum of these terms become constant. After reaching the equilibrium stage, the simulation was carried out with a microcanonical ensemble for 900 ps to obtain HCACF and VACF data.

\subsection{Thermal Conductivity Calculation}
\begin{figure}[H]
	\centering
	\includegraphics[width=12cm]{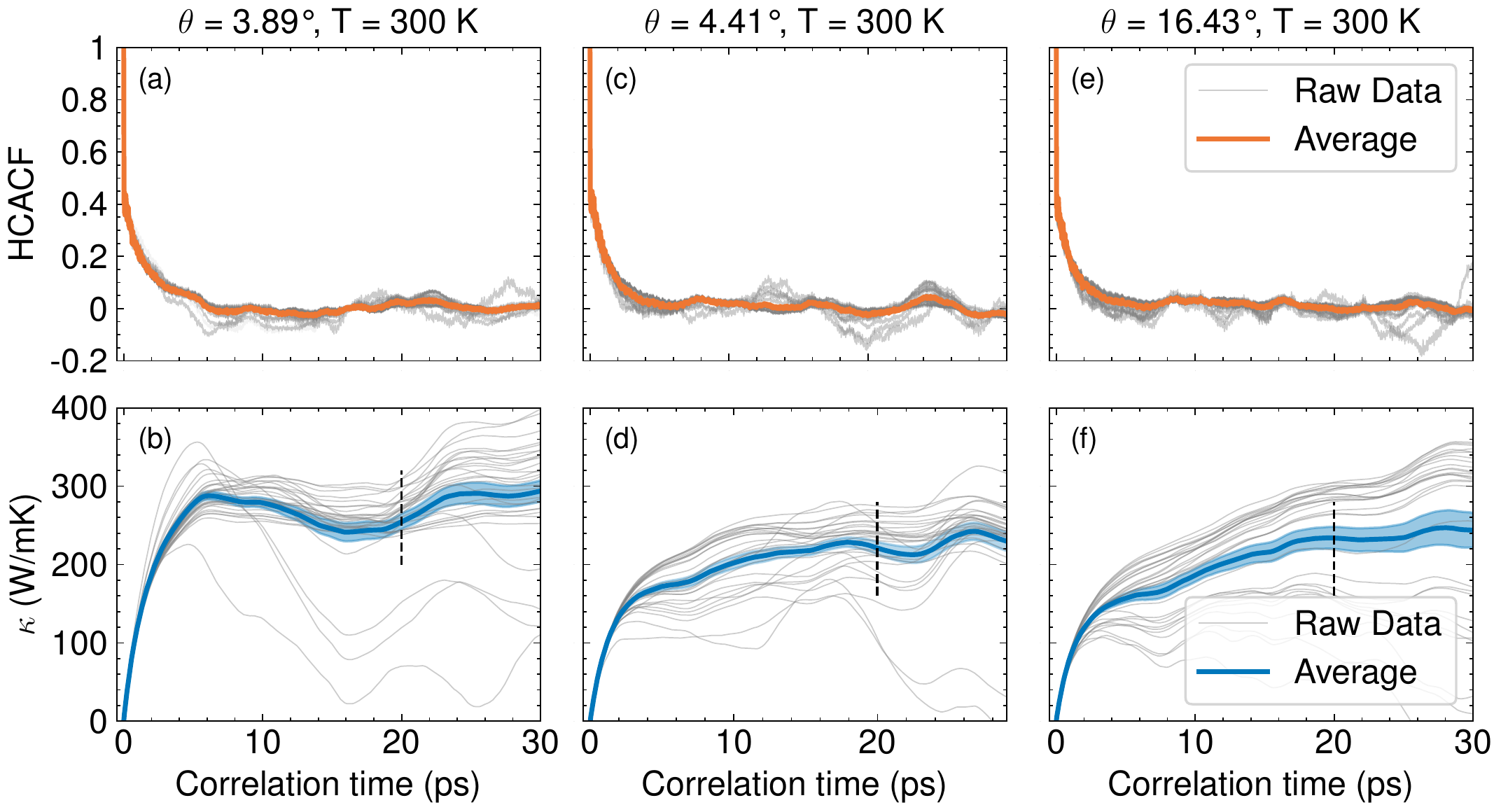}
	\caption{HCACF and Thermal Conductivity Evolution}
	\label{fig:hcacftc}
\end{figure}

Thermal conductivity calculation is carried out using the Green-Kubo theorem, by integrating the average heat flux autocorrelation function (HCACF). As a sample, Figure \ref{fig:hcacftc} shows the HCACF of tBLG with various twist angles at 300 K (a, c, e), and its integration over time (b, d, f). From the figure below, HCACF starts to decay to zero after 5 ps, and the integration of the HCACF becomes saturated, showing that the results are well-converged. Then, we average the integration data from 20 to 30 ps to accurately predict the thermal conductivities of our graphene structures.

\begin{figure}[H]
	\centering
	\includegraphics[width=7cm]{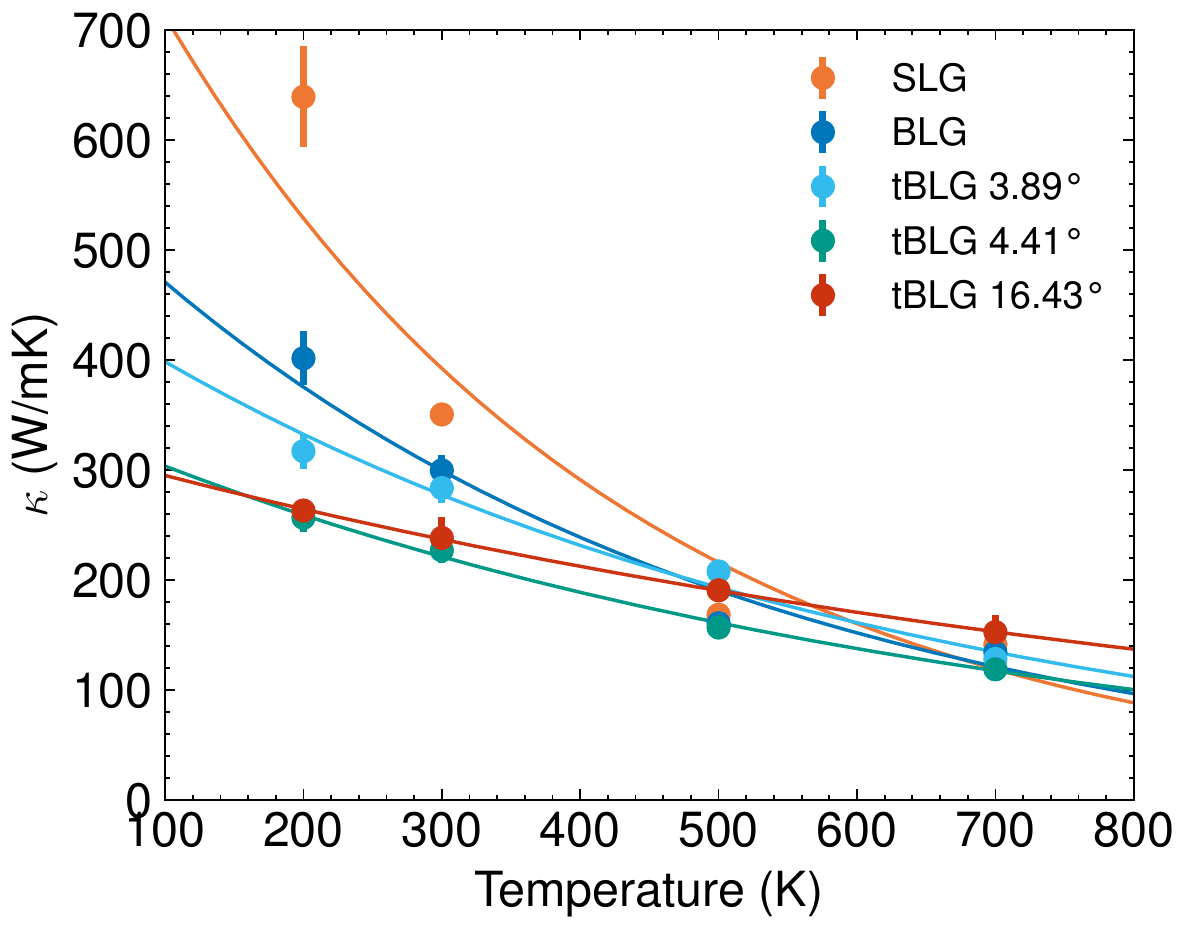}
	\caption{Thermal conductivity calculated from Molecular Dynamics}
	\label{fig:tcmd}
\end{figure}

The thermal conductivity of all systems we have calculated in this work is shown in Figure \ref{fig:tcmd}.

\noindent The thermal conductivity value from simulation results is represented by the dots, while the curve is the result of fitting the simulation result with an exponential function ($y=A e^{B x}$). The fitting is used to show the trend of thermal conductivity with respect to temperature and predict the value of thermal conductivity at temperatures that are not simulated.

The simulation shows that thermal conductivity decrease as temperature increases. This phenomenon can be explained theoretically and methodologically. In theory, phonons will be excited above the Debye temperature ($\mathrm{T}>\frac{h\omega}{k_{B}}$). By increasing the temperature, phonon-phonon scattering also increases, especially the Umklapp process (U-process) which is known as the dominant contributor to material's thermal resistivity.

\begin{figure}[H]
	\centering
	\footnotesize
	\subfloat{\includegraphics[width=4cm]{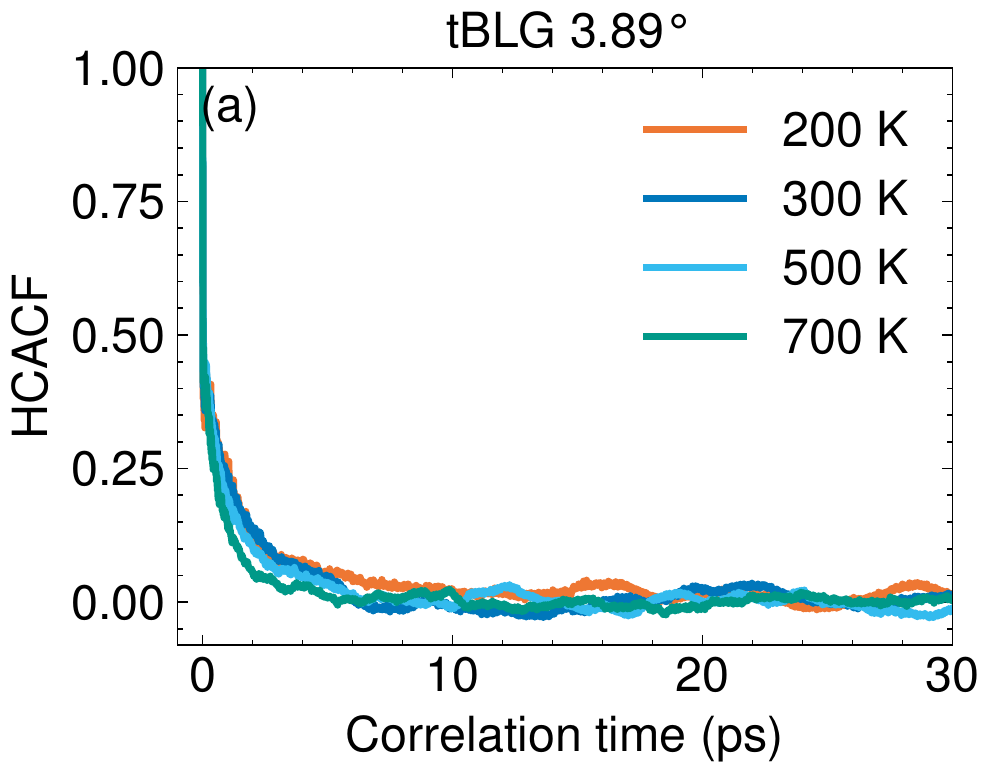}}
	\hspace{-1.5ex}
	\subfloat{\includegraphics[width=4cm]{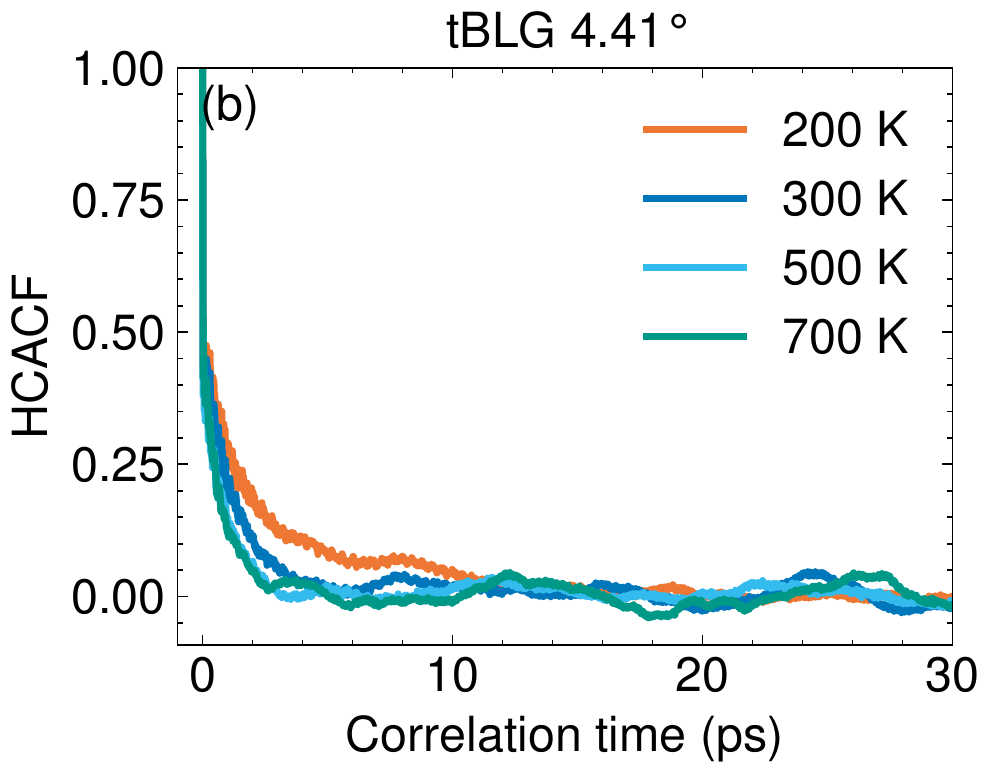}}
	\hspace{-1.5ex}
	\subfloat{\includegraphics[width=4cm]{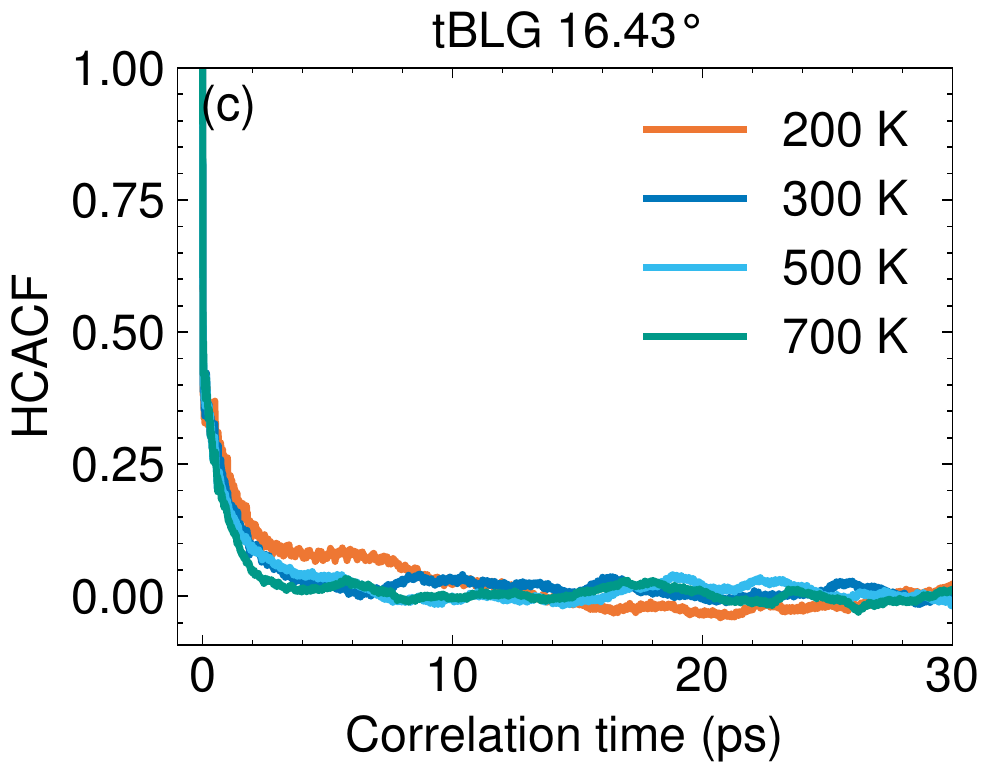}}
	\caption{HCACF on tBLG systems with different temperatures}
	\label{fig:decay}
\end{figure}

This phenomenon also can be explained by looking at how the Green-Kubo method works. The thermal conductivity value calculated from this method depends on how fast HCACF decay to zero. When the function has a positive value, the integration process will make the thermal conductivity value increase. As we see in Figure \ref{fig:hcacftc} (b, d, f), the integration graph is increasing at the early moment of simulation until it is saturated after HCACF decays and fluctuates around zero. It means when a function takes more time to decay, it also will take a longer time for the integration result to be saturated and keep increasing. As we can see in Figure \ref{fig:decay}, the HCACF decays faster in systems with higher temperatures.

\begin{figure}[H]
	\centering
	\footnotesize
	\subfloat{\includegraphics[width=5cm]{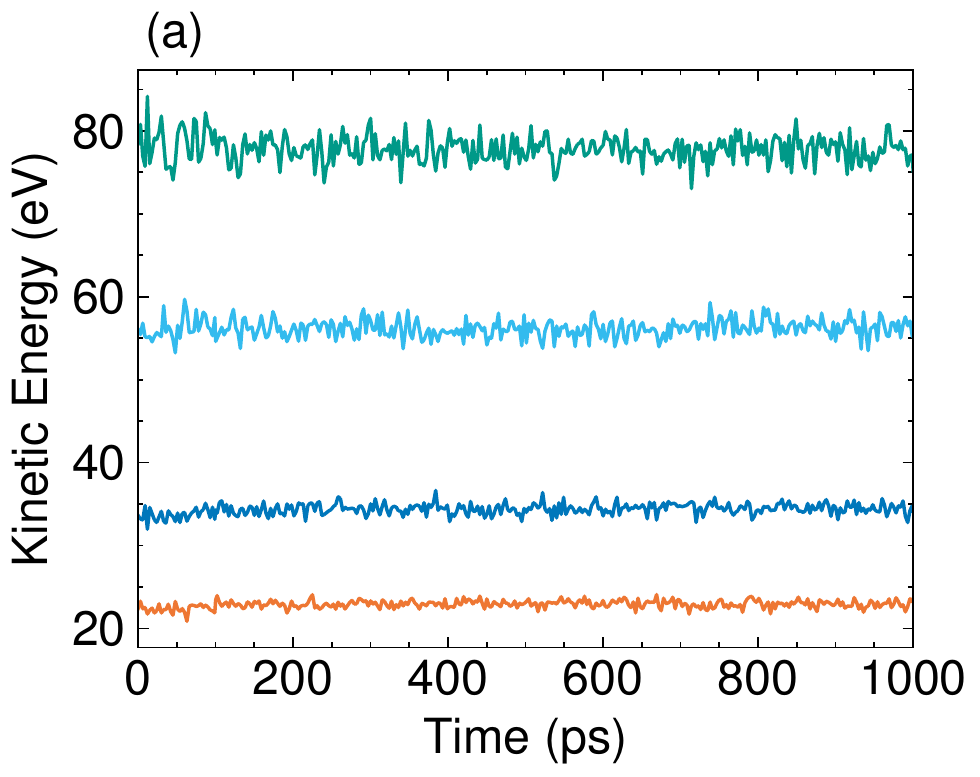}}
	\hspace{-1.5ex}
	\subfloat{\includegraphics[width=5.5cm]{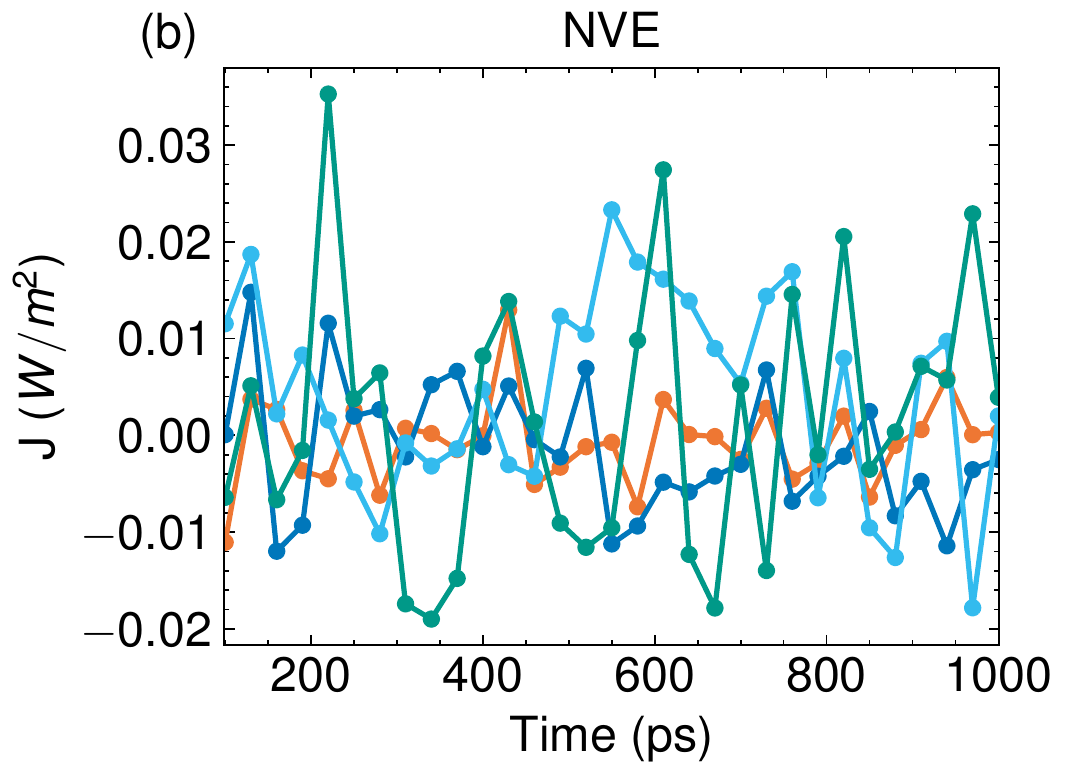}}
	\caption{Kinetic Energy and Heat Flux of tBLG 3.89\textdegree at various temperatures}
	\label{fig:kineticj}
\end{figure}

Figure \ref{fig:kineticj} shows how kinetic energy (a) and heat flux (b) of tBLG 3.89\textdegree in different temperature changes over time.

\noindent When the temperature of the system is increased, the kinetic energy of the system also increases, making the atoms become more actively moving and keep changing the system's states. As we can see, the system with a low temperature is persevering its state indicates by small displacements in the heat flux oscillation curve. The system with a high temperature has larger fluctuations, so when we try to find the correlation between each heat flux value, the correlation coefficient will tend to be zero (no correlation). Therefore, in classical molecular dynamics, the system with high temperature will have HCACF that decays faster and has smaller thermal conductivity compared to the system with low temperature.

\begin{figure}[H]
	\centering
	\includegraphics[width=10cm]{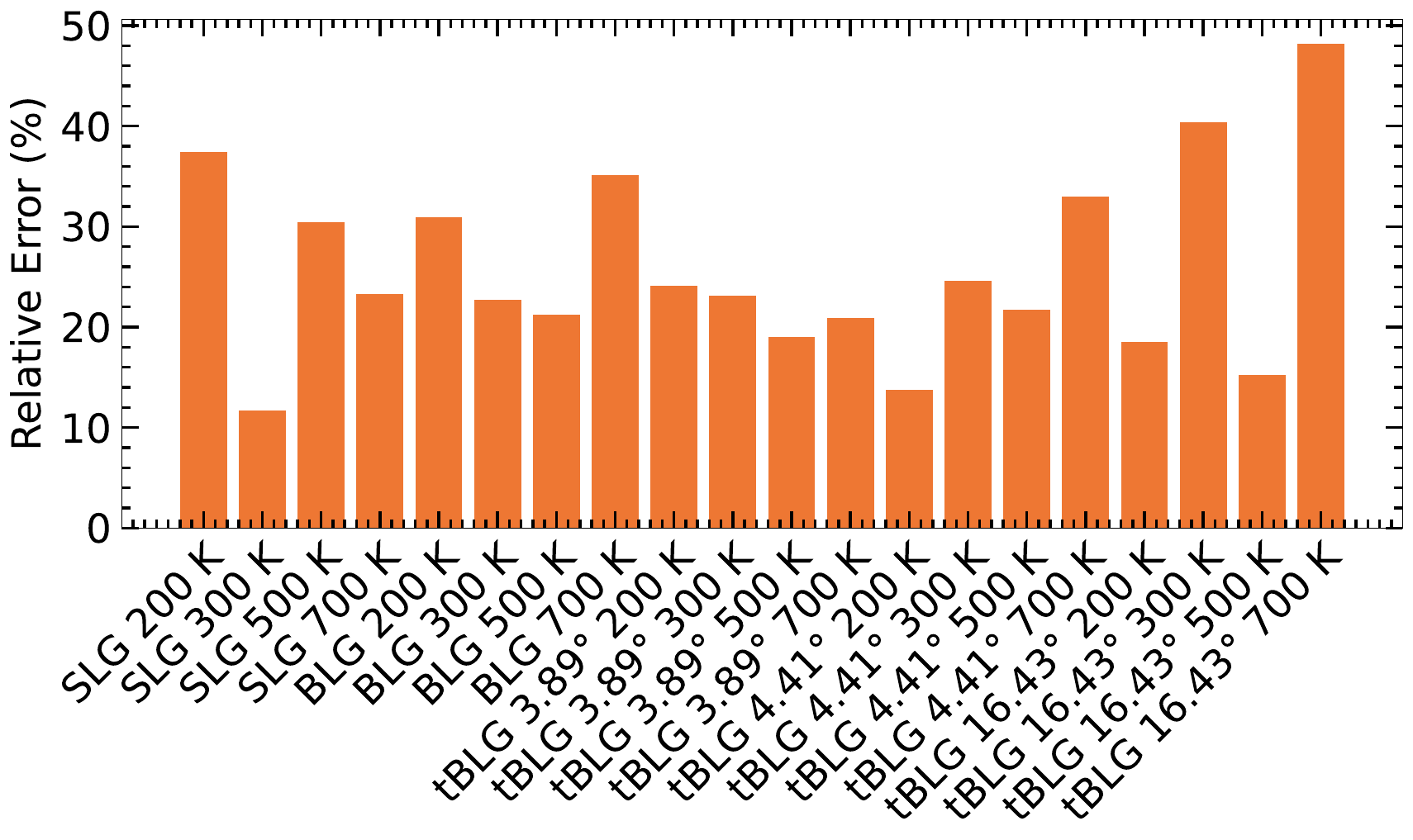}
	\caption{\centering Standard Error of Molecular Dynamics Simulation Results}
	\label{fig:err}
\end{figure}

The accuracy of methods and algorithms that have been used to perform this simulation was also evaluated by calculating the relative error, shown in the Figure \ref{fig:err}.

The highest relative error is 48.212\% from the simulation of tBLG 16.43\textdegree at 700 K, while the lowest relative error is 11.699\% from the simulation of SLG at 300 K, but each system has a relative error less than 50\%, pointing out that the simulation results are acceptable. Even though the simulation has converged, HCACF still fluctuates around zero, and how big this fluctuation is affecting the relative error produced from the simulation. The bigger the fluctuation, the greater the relative error value. For future research, one way to minimize the error is to extend the simulation and correlation time so that the HCACF fluctuation becomes smaller.

\subsection{Phonon Density of States}
Since heat transfer is reflected by phonon processes that occur in the material, exploring phonon characteristics like phonon density of state (PDOS) will give some knowledge on thermal conductivity. The PDOS is calculated from the Fourier transform of the velocity autocorrelation function (VACF) according to equation \ref{eq:PDOS}. The results of the PDOS calculations for each system at 300 K are shown in the following figure.

\begin{figure}[H]
	\centering
	\includegraphics[width=11cm]{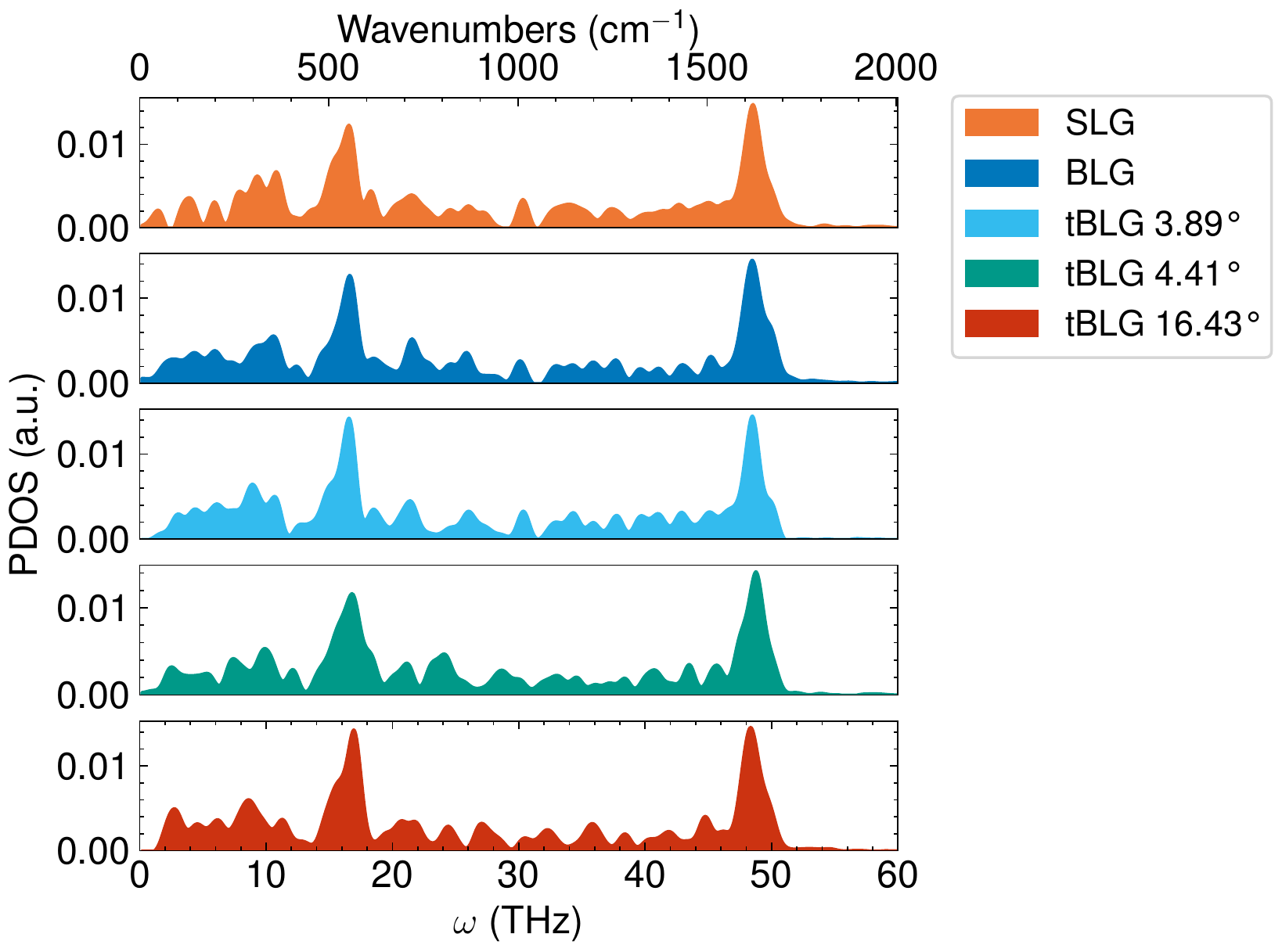}
	\caption{PDOS for each systems at 300 K}
	\label{fig:dosall}
\end{figure}

\noindent From the Figure \ref{fig:dosall}, the PDOS for the graphene system seems to have very small differences compared to each other. In this graphene system, there are 2 peaks around 15 THz and 48 THz. The 48 THz peaks are known as G-peak which is usually found in Raman spectroscopy measurements\cite{jorio2011raman, yang2015significant}.

\begin{figure}[ht!]
	\centering
	\begin{overpic}[width=12cm]{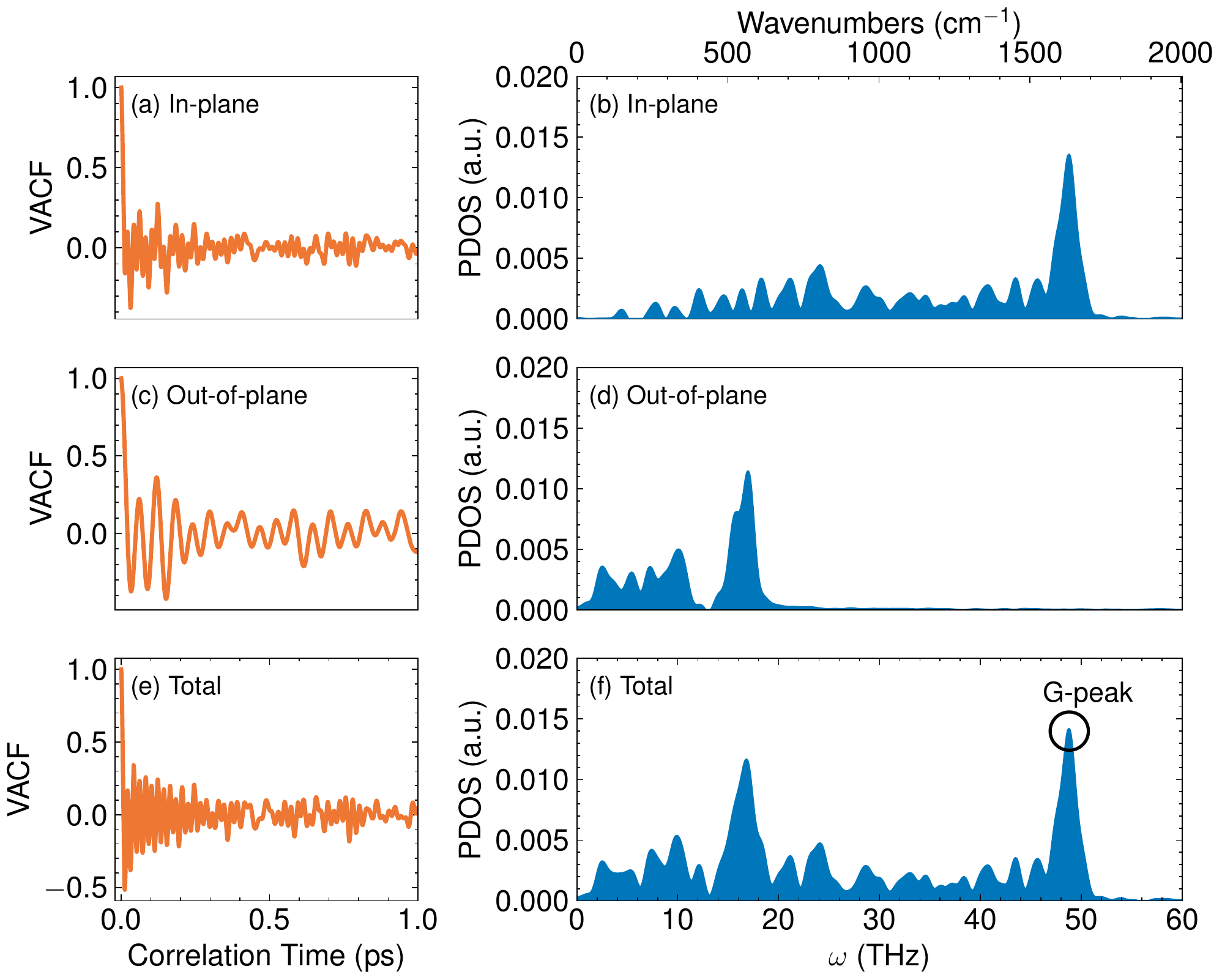}
		\put(74,14){\includegraphics[width=1.3cm]{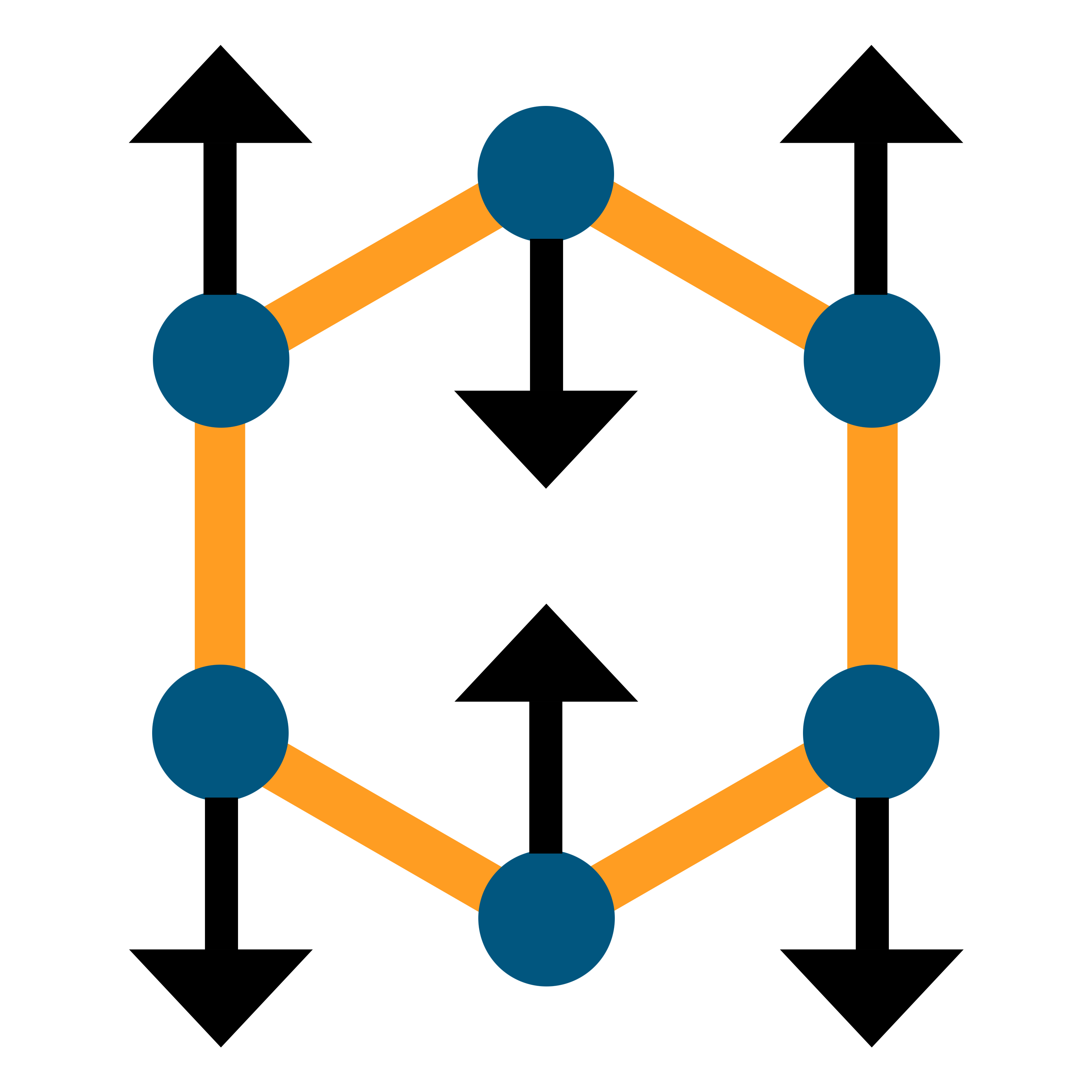}}
	\end{overpic}
	\caption[VACF and PDOS]{VACF and PDOS}
	\label{fig:vpdos}
\end{figure}

We also decomposed the PDOS based on the phonon modes. For example, the total and partial (in-plane and out-of-plane) contribution of PDOS for tBLG 3.89\textdegree is shown in the Figure \ref{fig:vpdos}. The x and y planes (in-plane mode) are dominated by phonons with high frequencies, known as optical phonons. Optical phonons have short mean free paths and are scattered more inside the nanostructures. Due to their high frequency and scattering location, these phonons have big contributions to phonon-phonon anharmonic interaction (U-process) and cause a decrease in thermal conductivity, especially in high temperatures. While the z planes (out-of-plane mode) are dominated by phonons with low frequencies, known as acoustic phonons, which are scattered much more at boundaries and interfaces.

\subsection{Spesific Heat Capacity and Quantum Correction}
\begin{figure}[H]
	\centering
	\footnotesize
	\subfloat{\includegraphics[width=6.0cm]{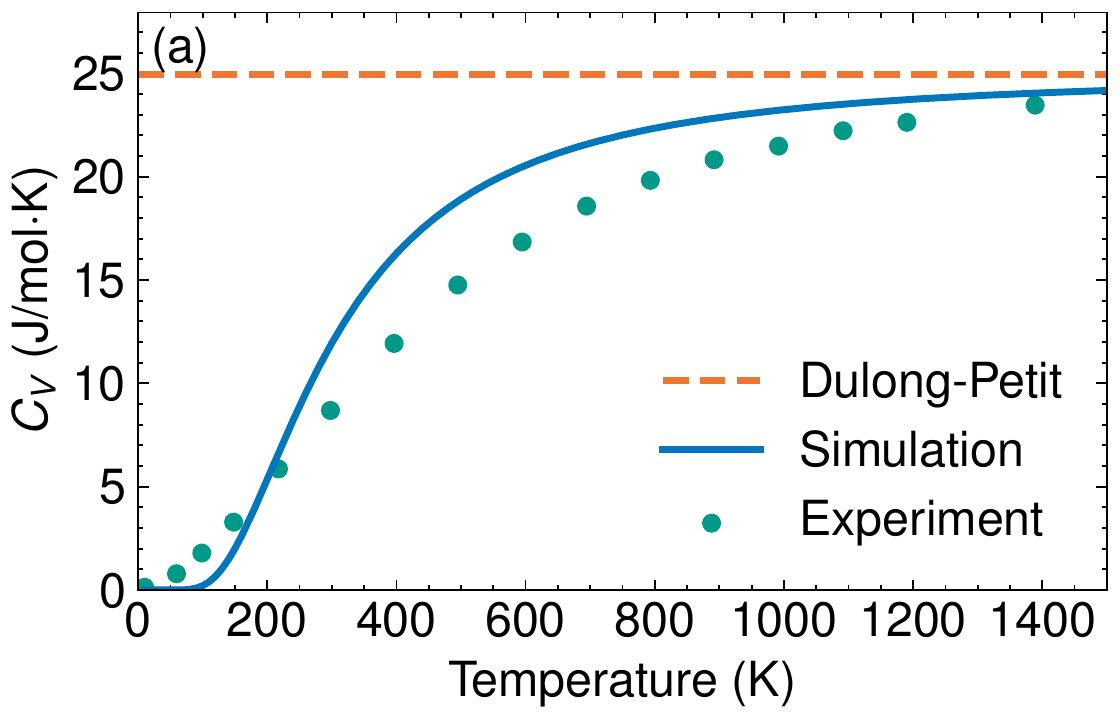}}
	\hspace{1.5ex}
	\subfloat{\includegraphics[width=5.2cm]{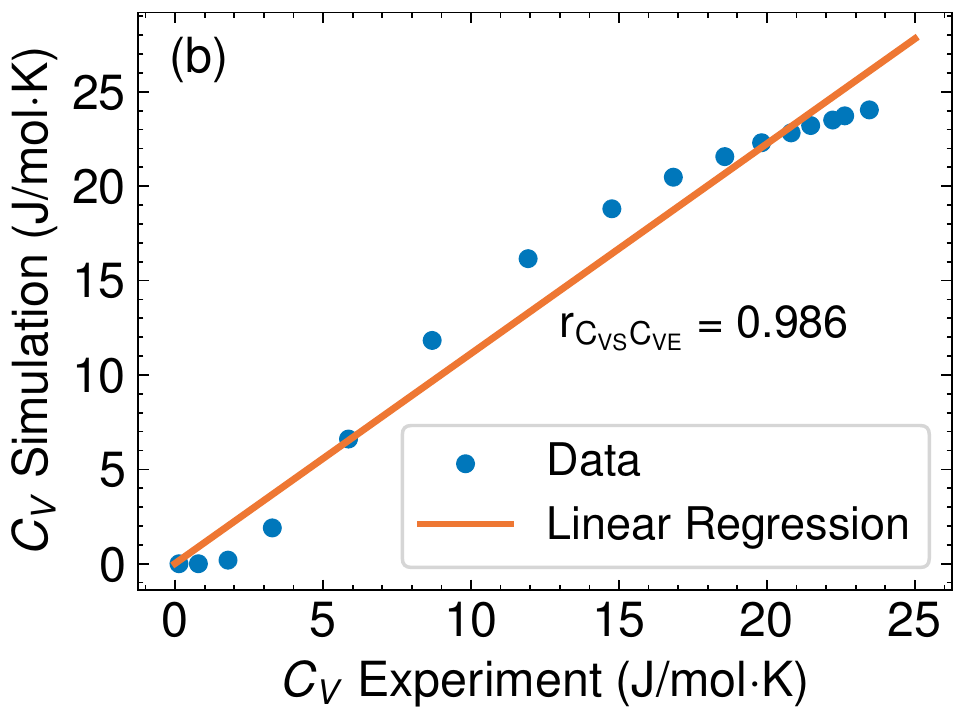}}
	\caption{Spesific Heat Capacity}
	\label{fig:cv}
\end{figure}

The specific heat capacity ($C_V$) of tBLG is calculated using Eq. \ref{eq:QCF} and shown by Figure \ref{fig:cv}.

\noindent The solid blue line is the simulation result of this work, the green circle is the experimental data from Pop\cite{pop2012thermal} and the dashed orange line is heat capacity based on the classical Dulong–Petit model. The simulation result shows that the $C_V$ is increasing as the temperature also increases, until when the temperature exceeds the Debye temperature ($\Theta_D$) the heat capacitance starts saturated on $3N_Ak_B = 3R \approx 25 J\cdot mol^{-1}K^{ -1}$ or known as Dulong-Petit limit, where $N_A$ is Avogadro's number, $k_B$ is Boltzman's constant, and $R$ is the ideal gas constant. Our simulation result agrees with the experimental data, with a correlation coefficient of 0.986.

The calculated $C_V$ value is used as quantum correction factor based on Eq. \ref{eq:QCF}. The quantum correction factors for various temperatures are shown in the following table.

\begin{table}[ht!]
	\centering
	\caption{Quantum Correction Factor}
	\begin{tabular}{|c|c|}
		\hline
		\textbf{Temperature (K)} & \textbf{Correction Factor} \\ \hline
		200                      & 0.2145                     \\ \hline
		300                      & 0.4790                     \\ \hline
		500                      & 0.7581                     \\ \hline
		700                      & 0.8666                     \\ \hline
	\end{tabular}
\end{table}

\begin{figure}[H]
	\centering
	\footnotesize
	\subfloat{\includegraphics[width=4cm]{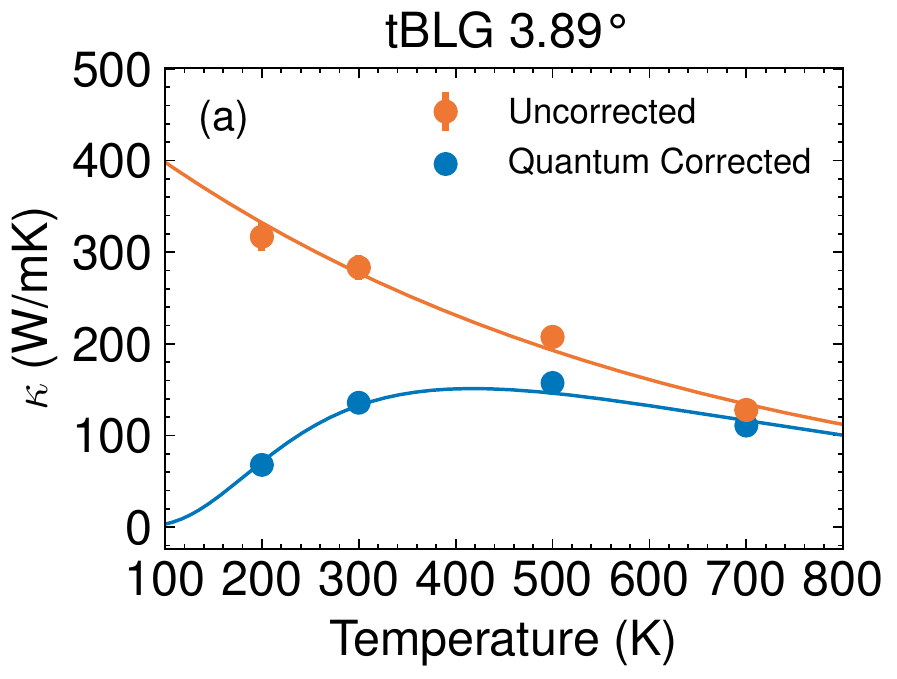}}
	\hspace{-1.5ex}
	\subfloat{\includegraphics[width=4cm]{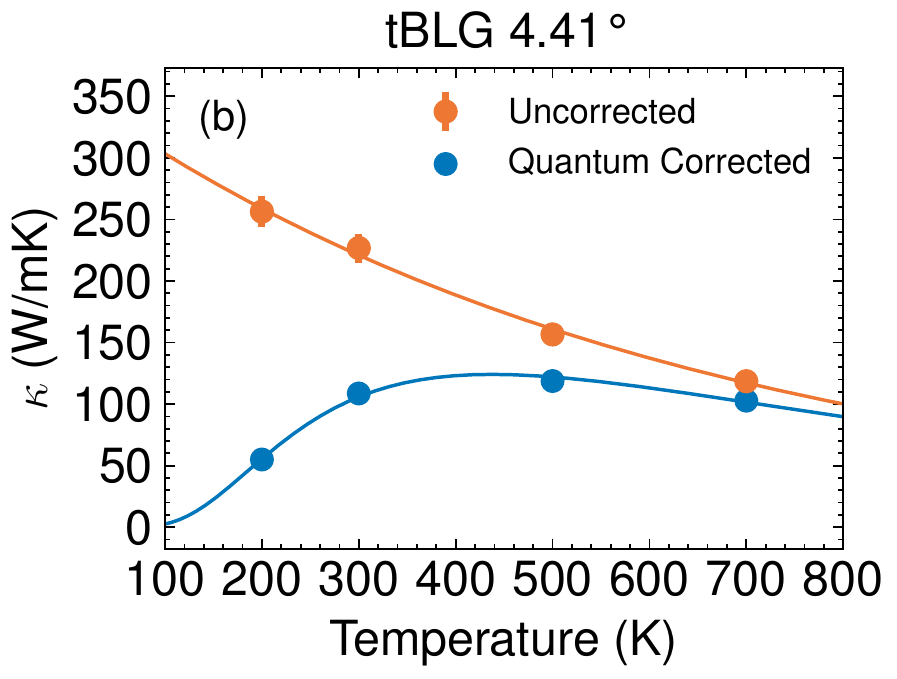}}
	\hspace{-1.5ex}
	\subfloat{\includegraphics[width=4cm]{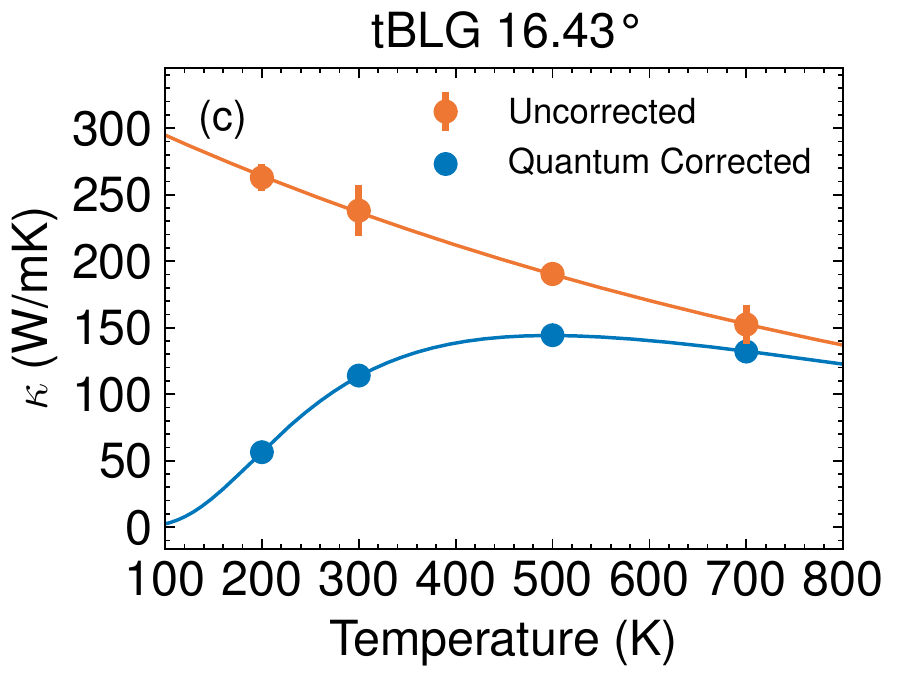}}
	\caption{Thermal Conductivity of tBLG from Classical Calculation and with Quantum Correction}
	\label{fig:qcorrection}
\end{figure}

The uncorrected thermal conductivity value from classical MD simulation and after correction is shown in Figure \ref{fig:qcorrection}.

\noindent In classical MD, the thermal conductivity decreases as the temperature is increasing. On the other hand, at low temperatures, the quantum-corrected thermal conductivity increases as temperature increases, until at a certain temperature the value of the thermal conductivity reaches its maximum and then decreases. The temperature value with the highest thermal conductivity is known as the Debye temperature ($\Theta_D$). The Debye temperature we get from our simulation for single layer graphene is around 340 K, which has a small difference from Hu's research\cite{hu2009molecular} at around 320 K. Meanwhile for bilayer graphene system, the $\Theta_D$ has a shift toward a higher temperature.

\begin{figure}[ht!]
	\centering
	\includegraphics[width=7cm]{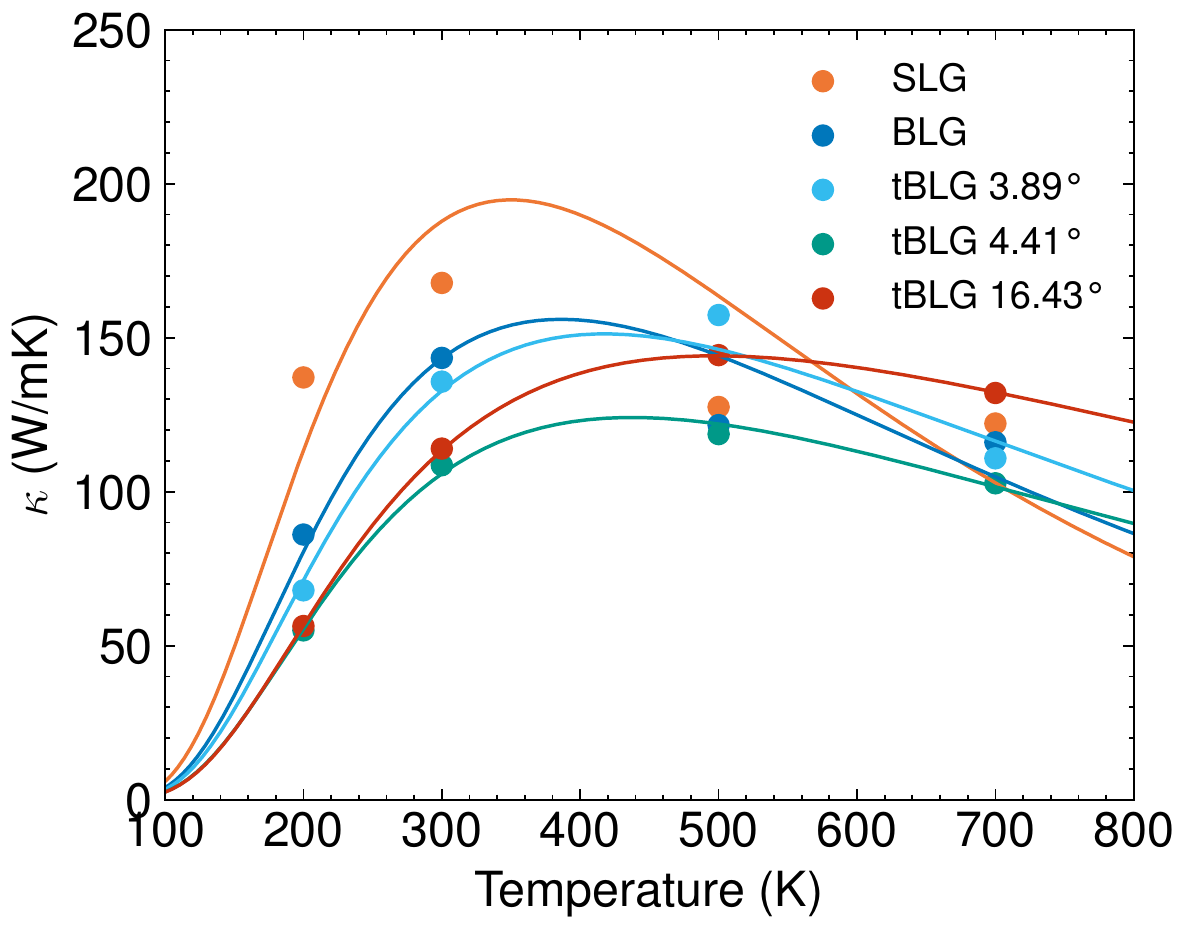}
	\caption{Thermal Conductivity with Quantum Correction}
	\label{fig:qtcsummarized}
\end{figure}

We summarized the quantum correction result of all graphene systems in Figure\ref{fig:qtcsummarized}. For all systems, initially at low temperature (T $<$ 100 K) $\kappa$ = 0. As the temperature increases, the initial increase in $\kappa$ is triggered by the effect of increasing heat capacity. When T $>$ $\Theta_D$, the heat capacitance value starts to become saturated according to the Dulong-Petit law, and the increasing phonon-phonon scattering gives the opportunity for a U-process to occur so that the thermal conductivity decreases.

\section{Conclusions}\label{sec4}
Investigations of thermal conductivity, phonon density of state, and specific heat capacity of twisted bilayer graphene can be carried out using Molecular Dynamics simulations based on the Green-Kubo theorem, and involving quantum correction. With quantum correction, the process of increasing thermal conductivity at low temperatures can be shown.

Simulation shows that stacking and twisting graphene layers causes the energy of the graphene system to change due to the shift of atomic positions relative to each other. These changes of energy on a microscopic scale contribute to microscopic quantities, including thermal conductivity. In this work, the highest $\kappa$ at around room temperature is owned by the tBLG with a twist angle of 3.89\textdegree followed by 16.43\textdegree and 4.41\textdegree. Despite possessing lower thermal conducting properties compared to single layer graphene, the tBLG is still can be categorized as a good conductor and suitable for thermal interface materials. The above results show the possibility of using tBLG in thermal management applications and thermoelectrics.

The specific heat capacity and phonon scattering affect the thermal conductivity of the material. At low temperatures, the thermal conductivity is increasing as the temperature increases, triggered by the specific heat capacity effect. After reaching Debye temperature ($\Theta_D$), the specific heat capacity effect becomes constant and the phonon scattering dominates. At high temperatures, thermal conductivity is decreasing due to the high phonon-phonon scattering in the so-called Umklapp process.

\section*{Acknowledgments}\label{sec5}
The authors wish to thank Prof. Camellia Panatarani, Prof. Ayi Bahtiar, and Nowo Riveli, Ph.D. for fruitful discussions during the final stages of this work.

\section*{Code availability}
The implementation code that support the findings of this study are openly available in GitHub at \href{https://github.com/dxid-dev/md-tblg}{github.com/dxid-dev/md-tblg}. This research has used LAMMPS\cite{LAMMPS}, VESTA\cite{momma2011vesta}, python and Atomic Simulation Environment (ASE)\cite{larsen2017atomic} \texttt{flatgraphene} module by Johnson Research Group at UIUC.

\section*{Disclosure statement}\label{sec6}
No potential conflict of interest was reported by the author.

\section*{Funding}
This research was financially supported by the Internal Grant of Universitas Padjadjaran, Indonesia.

\bibliography{sn-bibliography}
\end{document}